\documentclass[aps,prd,twocolumn,flushleft,amsmath,amssymb,nofootinbib]{revtex4}

\usepackage{graphicx}
\usepackage{dcolumn} 
\usepackage{bm}      

\newcommand{\bq}{\begin{equation}}
\newcommand{\eq}{\end{equation}} 
\newcommand{\ba}{\begin{eqnarray}}
\newcommand{\ea}{\end{eqnarray}}
\newcommand{\sinn}{{\rm sinn}}

\def\kms{{\rm km s}^{-1}}

\def\Mpc{{\rm Mpc}}

\begin{document}
\title{Two Component Model of Dark Energy} 
\author{Yan Gong}
\author{Xuelei Chen}
\affiliation{National Astronomical Observatories, Chinese Academy of
  Sciences, 20A Datun Rd, Chaoyang District, Beijing 100012, China}

\begin{abstract}
We consider the possibility that the dark energy is made up of two or more
independent components, each having a different equation of state. 
We fit the model with supernova and gamma-ray burst (GRB) data 
from resent observations, and use the Markov Chain Monte Carlo 
(MCMC) technique to estimate the allowed parameter regions.
We also use various model selection criteria to 
compare the two component model with the
$\rm \Lambda$CDM, one component dark energy model with static or variable
$w$ (XCDM), and with other multi-component models. 
We find that the two component models can give reasonably good fit to the 
current data. For some data sets, and depending somewhat on 
the model selection criteria, the two component model can give
better fit to 
the data than XCDM with static $w$ and XCDM with variable $w$ parameterized 
by $w = w_0 + w_az/(1+z)$.
\end{abstract}

\maketitle

\section{Introduction}

Observations have shown that more than 2/3 of the total cosmic density is 
made of an unknown ``dark energy'', which cause the expansion of the
Universe to accelerate \cite{Perlmutter99, Riess98}. 
The cosmological constant, with equation of state $w=-1$,
is the simplest form of dark energy, and it fits many observations
reasonably well. However, from a fundamental physics point of view, 
it suffers a fine-tuning problem of 122 order of
magnitude. Furthermore, it appears to be a great coincidence for its
density to be just comparable to the matter density today \cite{Carroll01}. 
Many alternative models of dark energy, e.g. 
quintessence\cite{Wetterich88, Ratra88,FJ97,CDS98, CLW98,Carroll98,ZMS99},
phantom\cite{Caldwell02}, quintom\cite{FWZ05}, 
K-essence\cite{Chiba00, Armendariz00}, 
Chaplygin gas\cite{Kamenshchik01, Bilic02, Bento02},
and so on, have been proposed, but at
present none of these is clearly superior than the cosmological
constant. Most of these models consists only a single component of dark
energy: either it is phenomenologically described by an equation of
state parameter $w: p=w\rho$ ($w$ can vary, and many different forms of
parameterization have been introduced), or by a dynamical field, although in
some cases two scalar fields are introduced. 

In the present paper, we consider dark energy models which 
consist of two independent components with equation of state $w_1$ and $w_2$
respectively.
This is motivated by the following considerisions. 
(1) If there is indeed unknown physical laws which give rise dark energy, 
it may well allow more than one form of it to be present. 
(2) Furthermore, if one describe dark energy by scalar fields, then they
do not necessarily have one single equation of state, especially in
the case with multiple fields, it is quite plausible that for each
field the corresponding equation of states are different. (3) Some
observations \cite{Riess04,Riess06} seem to favor models with $w<-1$ 
which necessitate negative kinetic terms (phantoms), or with 
variable equation of state in which $w$ vary accross -1 (quintoms), 
although in more recent 
observations \cite{Astier05,Wood07} these trends are 
weakened. One may wonder if such 
unusual behavior could be more naturally explained by considering two
dark energy components, with different equation of states, and
dominate at different redshifts. For example, in Ref.~\cite{Elizalde06}
two scalar fields were invoked to reproduce the phantom behavior.

We model the two independent dark energy components with equation
of state parameter $w_1, w_2$, which can either be fixed, or variable
with redshift. We use the Markov Chain Monte Carlo (MCMC) technique
to find the probablity distribution of the models in parameter space,
which allows us to see how the two component dark energy model fares
with observation. The data we used are supernova type Ia (SN Ia) and GRB data, 
which is currently the most direct and powerful probe of 
dark energy. In this analysis we have not included other constraints
, esp. the cosmic microwave background (CMB), as we do not expect the 
fixed effective equation of state description used here is really 
good at very high redshifts.

To compare our model with single component dark matter models and see
if there is indeed any evidence for two component of dark energy, we
calculate the best fit $\chi^2$, as well as the AIC, BIC and Bayesian 
evidence(BE). The models being compared include the $\rm \Lambda$CDM, XCDM
(phenomelogical model of single component dark energy described by an
equation of state), and our two component model $\rm X_1X_2CDM1$. 

We have used several different data sets to perform our fit, e.g.
the Gold04 \cite{Riess04} and Gold06 \cite{Riess06} data set, the 
SNLS \cite{Astier05} data set, and the ESSENCE data set\cite{Wood07}. 
There are some differeneces in the result. To avoid 
repetition, we presented the result of 
only one data set in detail, this is a combination of supernovae and
GRB data. The supernovae data
consists $182$ high-quality SN Ia samples \footnote{Although the
  number of sample coincides with that of Gold06, this is in fact a
  different data set.} selected 
from the Gold06 \cite{Riess06}, SNLS \cite{Astier05} and
ESSENCE \cite{Wood07} data sets, which includes 30 HST supernovae, 
47 SNLS supernovae from Gold06, 60 ESSENCE supernovae, and 45 nearby 
supernovae from WV07 \cite{Nesseris06}. Two 
different light-curve fitters, MLCS2k2 \cite{Riess96,Jha06}, and SALT
\cite{Guy05}, are used in these data sets.  
As they are consistent with each other \cite{Riess06, Wood07}, 
all of the SN Ia data used here are fitted 
using MLCS2k2 algorithm to avoid normalization \cite{Davis07}.
The GRB data provides a good complement to the SN Ia data
\cite{Ghirlanda04,Schaefer07,Liang05,Ghirlanda06,Dai04,Xu05,Firmani05,Hooper07},
as the GRB can be observed at much higher redshift, thus providing a
better baseline for detecting variation in the dark energy equation of
state. The GRB data set is constituted of 27 GRB samples in
Ref. \cite{Schaefer07}, which are generated with the $E_{peak}-E_{\gamma}$ 
correlation discovered by Ghirlanda et al.(2004) and is one of the tightest
correlation for GRB. We have not used all the $69$ GRB samples in that
paper, because there is larger uncertainty in the other correlations.
The method of calculating $\mu_{\rm GRB}$ is described in
Ref. \cite{Schaefer07}, but we only use the $E_{peak}-E_{\gamma}$
correlation in the paper. 
We marginalize $H_0$ at last as we did for the supernovae case.

The structure of this paper is as follows:  in \S 2 we present the
formalism and methods, including a brief summary of the  
MCMC approach for computing probablity distribution of the
parameters, and brief summary of model selection criteria. 
In \S 3, the results of our fit are shown.
In \S 4, we will compare a few models. Finally, we summarize our
results in \S 5.

\section{methods}

\subsection{Model}

We consider a model of two independent components of dark energy,
with static equation of state $w_1$ and $w_2$ respectively. The
expansion rate $H(z)$ is given by
\bq 
H^2(z) = H_0^2\ \Omega({\bf z;\theta}) ,
\eq
where, in the flat case:
\ba \label{fo:ome1} 
\Omega({\bf z;\theta}) &=& 
\Omega_{m_0} (1+z)^3 + \Omega_{x_1} (1+z)^{3(1+w_1)}\nonumber \\ 
&&+(1-\Omega_{m_0}-\Omega_{x_1}) (1+z)^{3(1+w_2)} 
\ea
and in the non-flat case:
\ba
\Omega({\bf z;\theta}) &=& (1-\Omega_{m_0}-\Omega_{x_1}-\Omega_{x_2}) (1+z)^2 
+\Omega_{m_0} (1+z)^3\nonumber\\
&& + \Omega_{x_1} (1+z)^{3(1+w_1)} + \Omega_{x_2} (1+z)^{3(1+w_2)} 
\ea 
Here ${\bf \theta}$ is the cosmological parameter set consisting of
\ba \label{fo:tc}
{\bf \theta} = \left\{\begin{array}{ll}
(\Omega_{m_0}, \Omega_{x_1}, w_1, w_2) & \textrm{in flat case}\\
(\Omega_{m_0}, \Omega_{x_1}, \Omega_{x_2}, w_1, w_2) 
& \textrm{in non-flat case}\end{array}\right.
\ea
The luminosity distance can be written as
\bq \label{fo:dl} d_L(z;{\bf \theta}) = (1+z)|\Omega_k|^{-\frac{1}{2}}\sinn\Big\{|\Omega_k|^\frac{1}{2}\times\int_{0}^{z}\frac{cdz'}{H(z')}\Big\}, \eq
where $\sinn(x)=\sinh(x),x,\sin(x)$ for open, flat and close geometries respectively.

\subsection{Analysis}

Given a cosmological model defined by $n$ parameters 
${\bf\theta} = (\theta_1,\ldots,\theta_n )$, and a data set consist
 of $N$ quantities ${\bf d} = (d_1,\ldots,d_N)$ with 
Gaussian-distributed errors ${\bf \sigma_d} =
(\sigma_1,\ldots,\sigma_N)$, 
the likelyhood function can be written as:
\bq \label{eq:likelihood}
{\mathcal{L}}({\bf d |\theta}) = 
\frac{1}{\sqrt{2\pi}{\bf\sigma_d}}e^{-\frac{1}{2}{\bf\chi}^2} ,
\eq
where for un-correlated data points, 
\bq \chi^2({\bf \theta}) = \sum_{i=1}^{N}\frac{(d^{obs}_i-d^{th}_i)^2}{\sigma_i^2}.\eq
If both SN Ia data and GRB data are used, we have
\bq  \chi^2 = \chi^2_{\rm SN_{sel}} + \chi^2_{\rm GRB} .\eq

For the SN Ia data, 
\bq \label{eq:chisq} \chi^2_{\rm SN}({\bf \theta}) = 
\sum_{i=1}^{N}\frac{(\mu_{obs}(z_i)-\mu_{th}(z_i))^2}{\sigma_i^2},\eq
where the theoretical value of distance modulus $\mu_{th}(z_i)$ is
given by
\ba \label{eq:mut} \mu_{th}(z_i) & = & 5\log_{10}d_L(z_i) + 25{} \nonumber \\
                  & = & 5\log_{10}D_L(z_i)-5\log_{10}h_0+42.38, \ea
and 
\bq D_L(z) = \frac{H_0}{c}\times d_L(z) .\eq

The Hubble constant $H_0 = 100 h_0 \kms\Mpc^{-1}$ may be marginalized
over as a nuisance parameters, as described in \cite{Nesseris05}, and
we would obtain 
\bq \chi^2_{\rm SN_{sel}}({\bf \theta}) = 
A({\bf \theta})\;- \;\frac{B^2({\bf \theta})}{C} ,\eq
where
   \ba   A({\bf \theta}) & = & \sum_{i=1}^{N}\frac{\Big(\mu_{obs}(z_i)
-5\log_{10}(D_L(z_i))\Big)^2}{\sigma_i^2}\nonumber \\
         B({\bf \theta}) & = & \sum_{i=1}^{N}\frac{\mu_{obs}(z_i)
-5\log_{10}(D_L(z_i))}{\sigma_i^2}\nonumber \\
         C & = & \sum_{i=1}^{N}\frac{1}{\sigma_i^2}  \ea 
Alternatively, we can integrate over $H_0$, and be left with
an additional constant term $\ln(C/2\pi)$ 
which would not affect our results given a 
data set\cite{Nesseris05, Nesseris04,Goliath01}.

\subsection{MCMC}
Given an observational data set ${\bf d}$, the posterior distribution of a 
parameter set ${\bf \theta}$ is, according to the Bayes theorem,
\bq \label{eq:pd} {\bf p(\theta|d)} = 
\frac{\mathcal{L}({\bf d|\theta})\:{\bf p(\theta)}}
{\int {\mathcal{L}(\bf d|\theta)p(\theta)d\theta}} ,\eq
where ${\bf p(\theta)}$ is the prior probability distribution, and
${\mathcal{L}}({\bf d | \theta})$ is the likelihood for obtaining 
the data set ${\bf d}$ given the parameter set ${\bf \theta}$ and is
given by Eq.~\ref{eq:likelihood}. To obtain the posterior probability
distribution of the parameters, we employ the MCMC technique to 
generate random samples in the parameter space. This method has
several advantages over grid-based approach. Most importantly,
the computational time cost increases linearly with the number of 
parameters, so even for a large number of parameters the estimate can
be done within an acceptable computation time. Additionally,
since the MCMC method generate samples from the full posterior
distribution, it gives far more information than the marginalized 
distributions\cite{Lewis02,book1, book2},
and dose not require a Gaussian distribution of the likelyhood\cite{Laurence06}. 

With the MCMC technique, a chain of sample points which is distributed
in the parameter space according to ${\bf p(\theta|d)}$ is 
generated by Monte Carlo. The Metropolis-Hastings algorithm with
uniform prior probability distribution is used to decide whether to
accept a new point into the chain by an acceptance probability:
\ba {\bf a}({\bf \theta_{n+1} | \theta_n}) & = & 
\min\bf \Bigg\{\frac{p(\theta_{n+1} | d)\;
q(\theta_n | \theta_{n+1})}{p(\theta_n|d)\;
q(\theta_{n+1}|\theta_n)}\ , 1\Bigg\}\nonumber \\ 
\nonumber\\& = & \min\bf \Bigg\{\frac{{\mathcal{L}}(d|\theta_{n+1})\;
q(\theta_n|\theta_{n+1})}{{\mathcal{L}}(d|\theta_n)\;
q(\theta_{n+1}|\theta_n)}\ , 1\Bigg\}  \ea    
where $\bf q(\theta_{n+1}|\theta_n)$ is the proposal density of
proposing a new point $\theta_{n+1}$ given a current point $\theta_n$
in the chain. If ${\bf a}= 1$, the new point $\theta_{n+1}$ is 
accepted; otherwise, the new point is accepted with probability
$\bf a$. The trials are repeated until a new point is accepted, and then  
we set $\theta_n = \theta_{n+1}$. In our computation, we set a uniform 
Gaussian-distributed proposal density for every point which is
independent of the position on the chain, so that 
$\bf q(\theta_{n+1}|\theta_n)$ and $\bf q(\theta_n|\theta_{n+1})$ are 
cancelled, we then have
\bq {\bf a}({\bf \theta_{n+1}|\theta_n}) = 
\min\Bigg\{\frac{{\mathcal{L}}\bf(d|\theta_{n+1})}{{\mathcal{L}}\bf(d|\theta_n)}\
, 1\Bigg\}  .\eq

We assume uniform prior for the parameters within the given ranges. We
have made tests with various ranges, for the results presented here we
have adopted the following:
$\Omega_{m_0} \in (0, 1 )$,  $\Omega_{x_1} \in (0, 2)$, 
$\Omega_{x_2} \in (0,2)$, $w_1 \in (-20, 5)$ and $w_2 \in (-20,5)$. 
We have considered also the special case of 
flat geometry case, for which
$\Omega_{x_1} \in (0, 1-\Omega_{m_0})$ is set, so that 
$\Omega_{x_2} = 1-\Omega_{m_0}-\Omega_{x_1}\ge 0$, assuming the 
energy density of all the components are non-negative.
   
Although the chains will eventually ``burn in'' even given a random
position in the parameter-space,  in practice we have chosen 
initial points near
the maximum likelihood point so that computational time 
is saved\cite{Eriksen05}.
   
   The proposal density have an important effect for good convergence 
and mixing of a chain, and the step provided by proposal density should 
not be either too large or small that may lead to slow convergence 
or mixing \cite{Lewis02, CosmoMC, Eriksen05, Christensen00, Doran04,
  Tegmark04, Hajian06, Hanson01, Verde03}. The method we use to
determine the proposal density is a adaptive step size Gaussian sampler
described in 
Ref. \cite{Doran04}. We use the criterion proposed by 
Gelman and Rubin (1992) to test convergence and mixing of the 
chains, and after that we freeze in the the proposal 
density \cite{Doran04, CMBeasy}. 
  
   We generate six chains for each case we study, and about seventy 
thousands points are sampled in each chain. After the burn-in process 
and thinning the chains,  we merge them into one chain which consists of 
about $10000$ points used to generate the probability distribution 
of the parameters. As the two dark energy components are completely
symmetric for Monte Carlo, when plotting the data 
we have also made ``mirrow'' points for
each point on the chain ($(\Omega_1,w_1) \leftrightarrow (\Omega_2,w_2$).

\subsection{Model selection criteria} 

We shall compare our two component models with one dark energy
component models. As the number of parameters differ, the simple
$\chi^2$ statistic is not always effective. We therefore consider 
several model selection criteria introduced in the 
literature \cite{Liddle04,Liddle06,Kurek07}.
  
  The Akaike information criterion (AIC) \cite{Akaike74} is
defined as
  \bq {\rm AIC} = -2\,\ln\mathcal{L}_{max} + 2\,k ,\eq
where $\mathcal{L}_{max}$ is the maximum likelihood and $k$ is the 
number of parameters in the model \cite{Liddle06, Biesiada07, Godlowski05}. The 
second term penalizes models with more parameters. 
However, the size of the data set is not considered
by the AIC, thus when we have very large number of data points, the
reduction on $\chi^2$ due to additional parameters would also be very
large, and one may still mis-select the model using the AIC criterion \cite{Liddle04}.

The Bayesian information criterion \cite{Schwarz78} can be written as
  \bq {\rm BIC} = -2\,\ln\mathcal{L}_{max} + k\,\ln N ,\eq
which includes the penalization of the number of data N.
Nevertheless, the BIC tends to over penalize the number of parameters,
given large number of data and is an approximation of the Bayesian evidence on some 
assumptions that may not be valid in practice, and it dose not 
take the full advantage of Bayesian technique \cite{Liddle04, Biesiada07, Kurek07}

The Bayesian evidence (BE) of a model $M$ takes the form 
   \bq {\rm BE} = \int {\mathcal{L}({\bf d|\theta}, M){\bf p}
(\mathbf{\theta}|M)d\mathbf{\theta}} ,\eq
which is just the denominator of equation(\ref{eq:pd}). The BE is the 
average of the likelihood of a model weighted by its prior in the parameter 
space. It automatically includes the penalties of the number of parameters
and data, so it is believed to be more direct, reasonable and unambiguous than 
the $\chi^2_{min}$ and ICs in model selection \cite{Liddle06,Liddle06a,John01,
Trotta05,Mukherjee06,Mukherjee05,Kunz06,Trotta07,Skilling}. 
The logarithm of BE can be used as a guide for comparing models (Jeffreys 1961),
and we choose the $\rm \Lambda$CDM as the referenced model for comparison:
$\Delta \ln({\rm BE}) = \ln({\rm BE})_{model}-\ln({\rm BE})_{\rm \Lambda CDM}$.
The strength of the evidence for the model is considered according to the
   numerical value of BE:
\ba
 \left\{\begin{array}{ll}
      \Delta \ln {\rm BE} < 1 & \quad \textrm{Weak}\\
       1 < \Delta \ln {\rm BE} < 2.5 & \quad \textrm{Significant}\\
       2.5 < \Delta \ln {\rm BE} < 5 & \quad \textrm{Strong to very strong}\\
       \Delta \ln {\rm BE} > 5 & \quad \textrm{Decisive}\end{array}\right.
\ea
We use the nested sampling algorithm 
to compute BE \cite{Mukherjee06,Skilling}, and the error is about $0.10$.

\section{Results}

\begin{figure}[htbp]
\includegraphics[scale = 0.35]{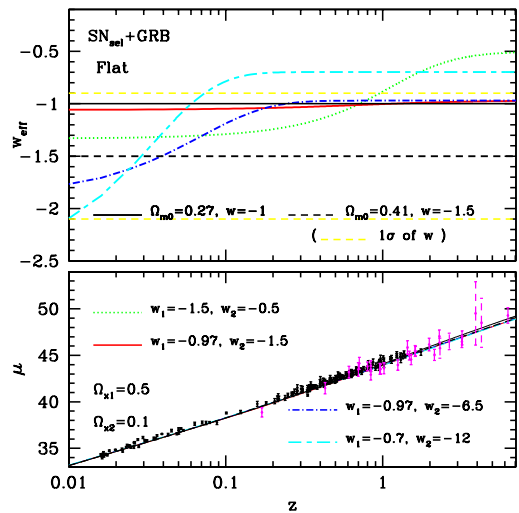} 
\caption{\label{fig:weff_mu} The $w_{eff}$ (upper panel) and luminosity
distance modulus $\mu$ (lower panel) 
for four two component models, marked with red(solid), green (dot),
blue (short dash dot) and cyan (long dash dot) lines. 
All these models have flat geometry and
$\Omega_{x1}=0.5,\Omega_{x2}=0.1$, but with different $w_1, w_2$. 
These model gave good (but not the best) fit to the data 
amongst the two component models. 
For comparison, the $\Lambda$CDM model with 
$\Omega_{m0}=0.27$ and the one component model with
$\Omega_{m0}=0.41,w=-1.5$ are plotted. For the latter, we also plot
the $1-\sigma$ error on $w$ in yellow (dashed) line.
In the $\mu$ plot, we 
show the SN and GRB data points and error bars in black and pink respectively.}
\end{figure}

For two component dark energy models, one can define an effective
equation of state, 
\begin{equation}
w_{eff}(z) = \frac{w_1 \Omega_{x1}(z)+w_2 \Omega_{x2}(z)}{\Omega_{x1}(z)+\Omega_{x2}(z)}.
\end{equation}
In Fig.~\ref{fig:weff_mu} we show the distance modulus $\mu$ and
equation of state $w_{eff}$ for a few examples 
of two component dark energy model, togather with the $\Lambda$CDM
model and an one-component dark energy model for reference. 
As can be seen from these examples,
the phantom divide line ($w=-1$) is easily crossed, which is not unexpected
if one component has $w>-1$ and another has $w<-1$.
The $w_{eff}$ and $\mu$ curves are very smooth, and 
the the two component model can fit the current data pretty well. 
Indeed, from these curves we may take note of the following character
of the two component model $\rm X_1X_2CDM1$: the effective dark energy
equation of state approaches that of a constant value at high
redshift, then it make transition to another value at some lower
redshift. $\rm XCDM2$, the one component dark energy model with 
variable equation of state with the usual parametrizations, 
can not realize this easily.
Inspecting the fit to the distance moduli curve, 
it is not easy to distinguish between these models from 
each other, or from the $\Lambda$CDM model and the one component dark
energy model, despite their apparant difference in the
$w_{eff}(z)$. The error on $w$ derived from current data is still too
large (the $1-\sigma$ error on $w$ for the one component model is plotted as
yellow dashed lines in Fig.~\ref{fig:weff_mu}).

\begin{figure}[tbp]
\includegraphics[scale = 0.35]{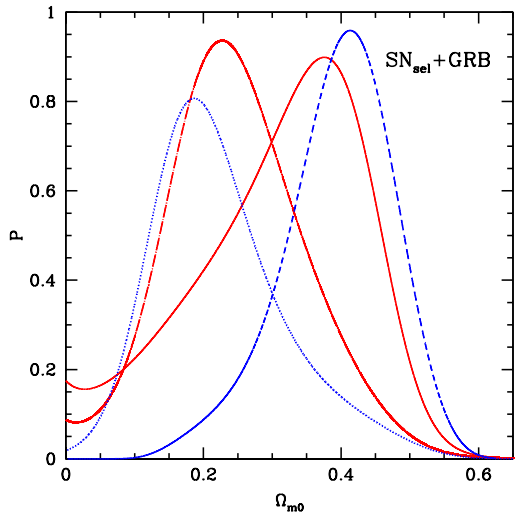}
\caption{\label{fig:m0_sg} The one dimensional PDF of $\Omega_{m_0}$.
The red solid line is for case of flat geometry, 
the red dot-dashed line for generic geometry. 
As a comparison, the PDF of one component model with static $w$ are shown,
and the blue dashed line is for flat geometry, the blue dotted line for generic geometry.}
\end{figure}

We now investigate the fit of the two component models more
quantatively, with MCMC simulation. 
We have performed our fit with several different combinations of data
sets. We found that the general features of the results are all very similar.
Here we present only one data set of $\rm SN_{sel}+GRB$ in detail.
We plot the PDF of $\Omega_{m0}$ in Fig.~\ref{fig:m0_sg}. 
The PDF of $\Omega_{m0}$ peaks
at 0.37 (0.22) for flat (generic) geometry. This value is typical for 
SN Ia only fit, but greater than
those based on the combined fit to SN Ia, CMB and LSS data. As we
have mentioned in the introduction, the purpose of this paper is to 
study the effect of two component dark energy on low redshift SN Ia
fits, we do not include CMB data. There is some difference in the best
fit (maximal point of the PDF curve) value of $\Omega_{m0}$ in the one
and two component models, but the difference is not too large.

\begin{figure}[htp]
\includegraphics[scale=0.35]{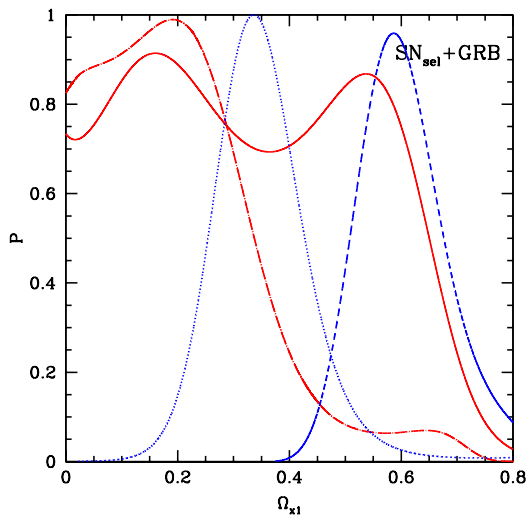}
\caption{\label{fig:x1_sg} The one dimensional PDF of $\Omega_{x1}$.
Red solid line: flat; red dot-dashed line: generic. For one component model
with static $w$, blue dashed line: flat; blue dotted line: generic.
}
\end{figure}

The PDF of one of the dark energy component, 
$\Omega_{x1}$ is plotted in Fig.~\ref{fig:x1_sg}. For reference, we
have also plotted the PDF of dark energy density in 
one-component dark energy models. Let us 
consider the flat case first. The PDF has a high plateau at
$\Omega_{x1}<0.65$ but declines sharply above that. As we have
$\Omega_{m0}+\Omega_{x1}<1$ and $\Omega_{m0}>0$, this result is not
surprising. There are also relatively small peaks and two small
troughs on the high plateau: the peaks at 0.16 and 0.54, and troughs at
0.02 and 0.37. We shall study the nature of these in the following
paragraphs. However, the peaks and troughs are not too large: the
amplitude is about 1/10 of the average. For the non-flat model, there
is a single peak at 0.2, and the PDF declines above the peak, but has
a small low plateau between 0.45 to 0.65.

\begin{figure}[htbp]
\includegraphics[scale = 0.35]{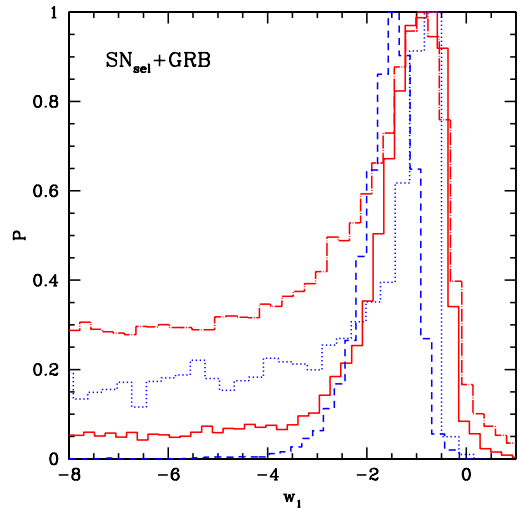} 
\caption{\label{fig:w1_sg} The PDF of equation of state parameter $w_1$,
and one component model are also shown.}
\end{figure}

\begin{figure}[htbp]
\includegraphics[scale = 0.35]{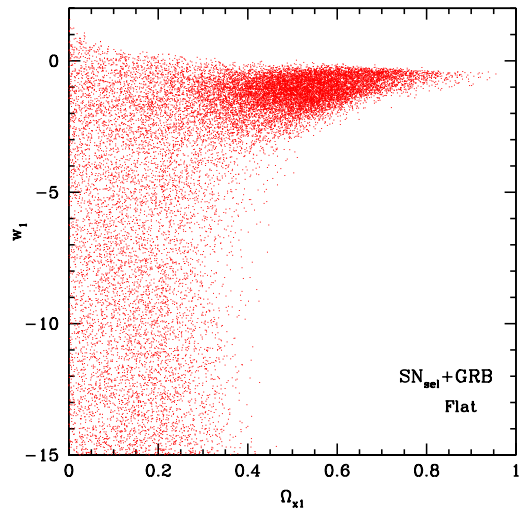} 
\caption{\label{fig:x1w1_sg} The point distribution of $\Omega_{x1}$-$w_1$.}
\end{figure}

In Fig.~\ref{fig:w1_sg} we plot the historgram for the distribution of
the equation of state parameter $w_1$. 
Again, for reference purpose we also plotted the
distribution for one component dark energy models.
For both the flat and general geometry case, the peak of the
distribution is near $w=-1$. However, for all these distributions there
are a long tails at $w_1 \ll -1$, for the non-flat case the tail
is more extended.

In Fig.~\ref{fig:x1w1_sg} we plot the distribution of $\Omega_{x1}$
vs. $w_1$. 
We can see that the density peak of the distribution is at 
$\Omega_{x1} \sim 0.55, w_1 \sim -1$. This corresponds well with the
peak in the PDF of $\Omega_{x1}$ and $w_1$. On the other hand, we note
that the distribution also spans to much negative value of $w_1$ with
small $\Omega_{x1}$: the addition of a small fraction of very negative
equation of state component (phantom) is allowed by the data. On the other
hand, for these deep phantom, large $\Omega_{x1}$ is not allowed,
leaving a conspicuous blank lower right half. The distribution do
extends to a ``pier'' at $w_1 =-1, \Omega_{x1}>0.6$. These are similar to 
$\rm \Lambda$CDM in the parameter space.
There is also some points on the corner with $w_1>0$.

\begin{figure}[htbp]
\includegraphics[scale = 0.35]{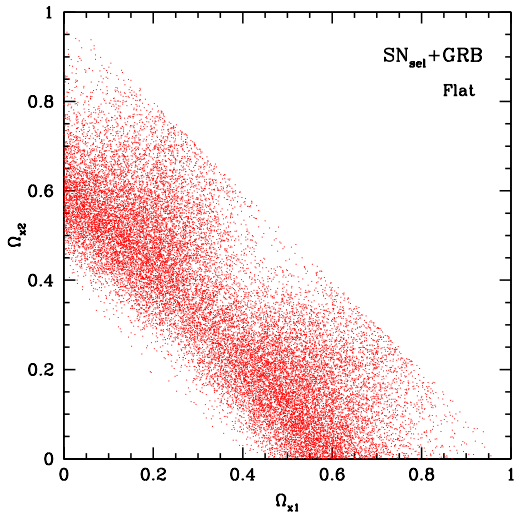} 
\caption{\label{fig:x1x2_sg} The point distribution of energy density for
the two components.}
\end{figure}

\begin{figure}[htbp]
\includegraphics[scale = 0.35]{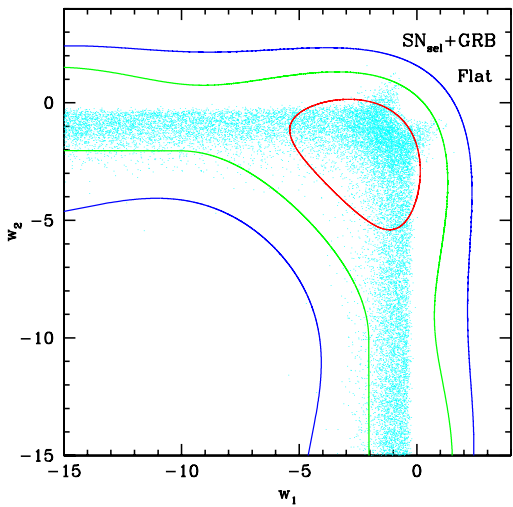} 
\caption{\label{fig:w1w2_sg} The point distribution and contour map
of equation of state for the two components.
The $1\sigma(68.3\%)$, $2\sigma(95.5\%)$ and $3\sigma(99.7\%)$ 
confidence levels are marked by red, green and blue solid lines respectively.}
\end{figure}

In the above we have looked at the distribution of one 
component in our two components model. Now we can also examine how the
two components are related. The scatter plot of $\Omega_{x1} \Omega_{x2}$
is shown in Fig.~\ref{fig:x1x2_sg}. Since
$\Omega_{x1}+\Omega_{x2} = 1-\Omega_{m0}$, it is very nature to see
that there is a linear anti-correlation between $\Omega_{x1}$ and
$\Omega_{x2}$. Apparantly, the region with
$0.4<\Omega_{x1}+\Omega_{x2}<0.7$ has greater density of points,
furthermore, in this region, there is a decrease of point density near
$\Omega_{x1} =\Omega_{x2}$.

We show the scatter plot and associated contours
for $w_1 w_2$ in Fig.~\ref{fig:w1w2_sg}. Near the $w_1=w_2$
diagonal the PDF is highest. Models in this region
is very similar to one component dark energy. As we
already know, the one component dark energy model fits reasonably well
the SN Ia data, so this result is not surprising. However, we note
that points far from the diagonal are also allowed, indeed the allowed
region is extends to very large negative $w$. These models have one
dark energy component which has $-2<w<0$, and another component whose $w$
can be smaller or much smaller than -1.

\begin{figure}[htbp]
\includegraphics[scale = 0.35]{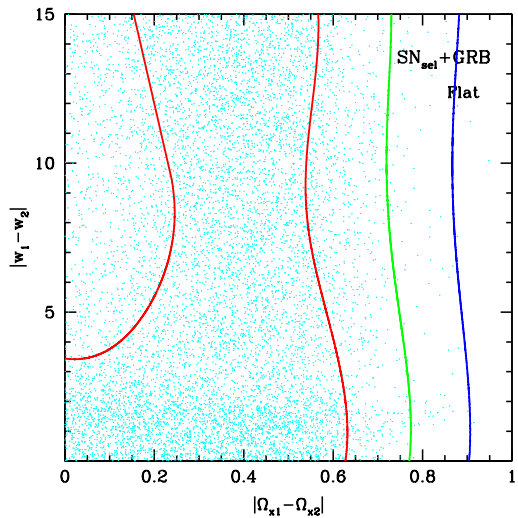} 
\caption{\label{fig:2d_sg} The difference of 
the two components.}
\end{figure}

Another way to see how the two components are related is to plot their
differences, as shown in Fig.~\ref{fig:2d_sg}, for
$|\Omega_{x1}-\Omega_{x2}|$ vs $|w_1-w_2|$. First, we note that the
points expands to very high value of $|w_1-w_2|$, showing that the two
components can be very different. Secondly, these high difference
points are widely distributed from $|\Omega_{x1}-\Omega_{x2}|=0$ to
$0.5$. This tells us that it is not the case that one single component
always dominates. Nevertheless, we can see that at 
$|\Omega_{x1}-\Omega_{x2}| \sim 0.4 $ the points are indeed somewhat more
frequent. 

\section{Model Comparisons}

In this section we compare the two component model considered in this
paper with several different models. We calculate their 
$\chi^2_{min}$, as well as several frequently used model selection criteria,
 namely AIC, BIC and BE.  The following models are 
considered: $\rm \Lambda$CDM, $\rm XCDM1$, $\rm XCDM2$, $\rm X_1X_2CDM1$,
 $\rm X_1X_2CDM2$, $\rm X_1X_2X_3CDM$ (see Table 1).  
$\rm XCDM1$ is an one component dark energy model with static $w$; $\rm XCDM2$ is an
one component model with variable $w$ parameterized by $w = w_0 +
w_a z/(1+z)$; $\rm X_1X_2CDM1$ is the two component model discussed extensively
in this paper. We also calculate the statistics of other two kind of 
multi-component models: 
$\rm X_1X_2CDM2$ is two components model with variable $w$ parameterized 
by $w_1 + w_{1a} z/(1+z)$ and $w_2 + w_{2a} z/(1+z)$; $\rm X_1X_2X_3CDM$ is 
a three component model with static $w$.

With different data set, the best fit result is also different. 
Here we consider several different data sets: (1) the selected supernovae
samples and GRB samples, for which detailed result is presented in the
last section; (2) the selected supernovae sample, but without GRB; (3)
182 supernovae known as the Gold06 sample (4) The SNLS sample, which
includes 115 supernovae sample published 
in Ref.~\cite{Astier05}, for this sample the parameters 
$\alpha$ and $\beta$ are joined into our
cosmological parameter set to estimate the statistics; (5) the 
ESSENCE data set, which consists of two parts: $60$ ESSENCE samples 
and $45$ nearby SN Ia published in Ref.~\cite{Wood07}.

Using the above data sets, we calculate the $\chi^2$ and other model selection
criteria. Our calculation is done for flat and non-flat geometric
prior seperately. The results are given in Table 2 
and Table 3. 

First we look at the flat case. It is clear that with additional
parameters, the $\chi^2$ can always be reduced for models with more
parameters, hence it not very useful in comparing models with different 
number of parameters\cite{evidence}. On the other hand, using different model selection
criteria, the conclusion we draw could differ. The BIC strongly
penalizes models with more parameters, so models with less parameters 
almost always win. For the AIC, the results varies with
different data set. For the selected SNIa data 
set $\rm SN_{sel}$ (which are selected for the
best data) and $\rm SN_{sel}+GRB$ data set, 
the $\rm X_1X_2CDM1$ fares as well as $\rm XCDM1$
(the one component dark energy model with fixed $w$), and  
better than the $\rm XCDM2$ (the one component dark energy model with
variable $w$). In the ESSENCE data set, the $\rm X_1X_2CDM1$ fits
almost as well as $\rm XCDM1$ and better than $\rm XCDM2$.
In the Gold 06 and SNLS data sets, however, the one component models
fit better than two component models.

Perhaps the most accurate model selection criterion considered here is
BE. Comparing to $\Lambda$CDM, 
for most data set the two component model 
$\rm X_1X_2CDM1$ is slightly disfavored using BE 
(except for the ESSENCE data set, for
which it is slightly favored). However, the evidence of this disfavorance is 
weak. In comparison, the one component dark energy model $\rm XCDM1$ and
$\rm XCDM2$ are significantly disfavored when compared with the $\rm
\Lambda$CDM. We also found that $\rm X_1X_2CDM2$ and $\rm
X_1X_2X_3CDM$ have comparable BE.

While we have reason from both theory (inflation) and observation
(CMB) to suppose that the geometry of observable universe is flat, for
generality we also consider non-flat geometry.
In the non-flat case, again with different data sets the result are
somewhat different. Generally, however, the data seems to be less
favorable for two component dark energy models. For example, with
$\rm SN_{sel}+GRB$ data set, the $\rm X_1X_2CDM1$ is significantly disfavored over 
$\rm \Lambda$CDM, which is similar to $\rm XCDM1$, while the $\rm XCDM2$ is only
slightly disfavored. However, $\rm X_1X_2CDM2$ and $\rm X_1X_2X_3CDM$ is slightly
favored. We suppose that in the non-flat model, the curvature term
behavior is degenerate with a special form of dark energy component,
which makes the model less favored by the model selection
criteria.

\section{Conclusion}

In this paper we considered phenomelogical multi-component
dark energy models, in particular a two component model
with static equation of state $w_1$ and $w_2$, and examined how good
can they fit the supernovae and GRB data. We have tested our models
with several recently published data sets. For most of the analysis, 
we have used a high quality data set with 182 SN Ia sample 
and $27$ GRB sample selected from resent 
observations. The MCMC technique is used in our cosmological parameter 
fitting, and to compare different models we have tried 
several frequently used model selection criteria.

The fitting results indicate that the two component models can fit 
the current data set fairly well. The effective equation of state 
of the dark energy can cross the phantom divide line, as would have
been expect for a model with one component $w>-1$ and another
$w<-1$.

We have compared the multi-component models with
the $\rm \Lambda$CDM model and one component models using several 
model selection criteria, including $\chi^2$, AIC, BIC, and BE.
The result varies for different choice of model selection criteria and
data set. The BIC almost always favor models with less parameters. For
AIC, we found different results for different data sets.
Since the Bayesian evidence (BE) is presently considered to be
better than either AIC or BIC, 
we have mainly focused on the results derived using BE. 
Using BE, we found that for flat geometry the multi-component models 
can fit as well as the $\rm \Lambda$CDM model, and for 
all of the data sets, which is worthy of note, the multi-component models
fit significantly ($\rm \Delta \ln(BE)\gtrsim1$) better than one 
component models.

The model selection technique has been applied to dark energy model
studies in several recent works. For the most recent data, the
$\Lambda$CDM model is favored over models with simply parametrized 
variable equation of state models such as 
XCDM1 and XCDM2 \cite{Serra07, Lazkoz07}. We have reached the same 
conclusion for these models. Because the data set and constraining
technique used in each case is different (e.g., in
Ref.~\cite{Serra07}, SNIa Gold 06 data set, togather with CMB shift
parameter and SDSS BAO result is used), it is difficult to make 
a full comparison. We note, however, that 
 for $\rm XCDM1$ and $\rm XCDM2$ they obtained 
$\rm \Delta \ln(BE) = -1.027, -1.118 $ 
(with respect to $\Lambda$CDM) respectively. 
We have not used the CMB and BAO data in our
calculation, but for the Gold 06 data set, we found $\rm \Delta
\ln(BE) = -1.60, -1.16$ respectively, which are of similar strength.

It is premature to say at this stage whether single or multiple
component dark energy model is favored, as the evidence is still
weak. Even if in the future we find that multi-component model is
preferred in a fit, we note that this is only a phenomelogical model
based on parametrazation. It is conceivable that dark energy of a
single physical origin could give rise to the appearance of multiple 
components with different equation of state. Nevertheless, our investigation
show that multiple component is allowed by current data. When one
discusses the properties of dark energy as inferred from observations,
this possibility should not be neglected.

\begin{acknowledgments}
Our MCMC chain computation was performed on the Supercomputing Center of 
the Chinese Academy of Sciences and the Shanghai Supercomputing
Center. This work is supported by
the National Science Foundation of China under the Distinguished Young
Scholar Grant 10525314, the Key Project Grant 10533010, by the
Chinese Academy of Sciences under grant KJCX3-SYW-N2, and by the
Ministry of Science and Technology under the national basic sciences
program (973) under grant 2007CB815401.   
\end{acknowledgments}

\begin{table*}[htbp]
\caption{H(z) of the models}
\begin{ruledtabular}
\begin{tabular}[t]{c||c|c}
\bf Model & \bf Flat & \bf Non-flat\\
\hline
$\rm \Lambda CDM$ & $H^2(z)=H^2_0[\Omega_{m0}(1+z)^3+(1-\Omega_{m0})]$ &
 $H^2(z)=H^2_0[\Omega_{m0}(1+z)^3+\Omega_{\Lambda 0}+(1-\Omega_{m0}-\Omega_{\Lambda 0})(1+z)^2]$\\
\hline
$\rm XCDM1$ & $H^2(z)=H^2_0[\Omega_{m0}(1+z)^3$ &
 $H^2(z)=H^2_0[\Omega_{m0}(1+z)^3+\Omega_x(1+z)^{3(1+w)}$ \\
 & $+(1-\Omega_{m0})(1+z)^{3(1+w)}]$ &  $+(1-\Omega_{m0}-\Omega_x)(1+z)^2]$\\
\hline
$\rm XCDM2$ & $H^2(z)=H^2_0[\Omega_{m0}(1+z)^3$ &
 $H^2(z)=H^2_0[\Omega_{m0}(1+z)^3+\Omega_x(1+z)^{3(1+w+w_a)}$ \\
 &$+(1-\Omega_{m0})(1+z)^{3(1+w+w_a)}e^{3w_a[1/(1+z)-1]}]$ &  $e^{3w_a[1/(1+z)-1]}+(1-\Omega_{m0}-\Omega_x)(1+z)^2]$\\
\hline
$\rm X_1X_2CDM1$ & $H^2(z)=H^2_0[\Omega_{m0} (1+z)^3 + \Omega_{x1} (1+z)^{3(1+w_1)}$ & $H^2(z)=H^2_0[\Omega_{m0} (1+z)^3 + \Omega_{x1} (1+z)^{3(1+w_1)}$\\
&$+(1-\Omega_{m0}-\Omega_{x1})(1+z)^{3(1+w_2)}]$ & $+\Omega_{x2}(1+z)^{3(1+w_2)}+(1-\Omega_{m0}-\Omega_{x1}-\Omega_{x2}) (1+z)^2]$\\  
\hline
 &$H^2(z)=H^2_0[\Omega_{m0}(1+z)^3$ & $H^2(z)=H^2_0[\Omega_{m0}(1+z)^3$ \\
$\rm X_1X_2CDM2$&$+\Omega_{x1}(1+z)^{3(1+w_1+w_{1a})}e^{3w_{1a}[1/(1+z)-1]}$&$+\Omega_{x1}(1+z)^{3(1+w_1+w_{1a})}e^{3w_{1a}[1/(1+z)-1]}$\\
 &$+(1-\Omega_{m0}-\Omega_{x1})(1+z)^{3(1+w_2+w_{2a})}$ & $+\Omega_{x2}(1+z)^{3(1+w_2+w_{2a})}e^{3w_{2a}[1/(1+z)-1]}$ \\
 &$e^{3w_{2a}[1/(1+z)-1]}]$& $+(1-\Omega_{m0}-\Omega_{x1}-\Omega_{x2})(1+z)^2]$ \\
\hline
& $H^2(z)=H^2_0[\Omega_{m_0} (1+z)^3 + \Omega_{x_1} (1+z)^{3(1+w_1)}$ & $H^2(z)=H^2_0[\Omega_{m_0} (1+z)^3 + \Omega_{x_1} (1+z)^{3(1+w_1)}$ \\
$\rm X_1X_2X_3CDM$ & $+\Omega_{x2} (1+z)^{3(1+w_2)}$ & $+\Omega_{x2} (1+z)^{3(1+w_2)} +\Omega_{x3}(1+z)^{3(1+w_3)}$\\
& $+(1-\Omega_{m0}-\Omega_{x1}-\Omega_{x2})(1+z)^{3(1+w_3)}]$ & $+(1-\Omega_{m0}-\Omega_{x1}-\Omega_{x2}-\Omega_{x3})(1+z)^2]$ \\

\end{tabular}
\end{ruledtabular}
\end{table*}

\begin{table*}[htbp]
\caption{Flat case}
\begin{ruledtabular}
\begin{tabular}[t]{c||c|c|c|c|c|c|c}
\bf Data & \bf Type & $\rm \Lambda CDM$ & $\rm XCDM1$ & $\rm XCDM2$ & $\rm X_1X_2CDM1$ & $\rm X_1X_2CDM2$ & $\rm X_1X_2X_3CDM$\\
\hline
 & $\chi^{2}_{min}$ & 158.749 & 156.584 & 156.405 & 156.453 & 155.052 & 156.430\\
$\rm Gold06$ & $\rm AIC$ & 160.749 & 160.584 & 162.405 & 164.453 & 167.052 & 168.430\\
(182) & $\rm BIC$ & 163.953 & 166.992 & 172.017 & 177.269 & 186.404 & 187.654\\
 & $\rm \Delta \ln(BE)$ & - & -1.60 & -1.16 & -0.68 & +0.04 & +0.15\\
\hline
 & $\chi^{2}_{min}$ & 110.033 & 110.998 & 111.008 & 110.704 & 110.110 & 109.707\\
$\rm SNLS$ & $\rm AIC$ & 117.033 & 118.998 & 121.008 & 122.704 & 126.110 & 125.707\\
(115) & $\rm BIC$ & 125.268 & 129.978 & 134.733 & 139.173 & 148.070 & 147.666\\
 & $\rm \Delta \ln(BE)$ & - & -1.26 & -1.34 & -0.36 & +0.66 & +0.64\\
\hline
 & $\chi^{2}_{min}$ & 104.531 & 103.459 & 103.391 & 99.469 & 98.150 & 97.311\\
$\rm ESSENCE$ & $\rm AIC$ & 106.531 & 107.459 & 109.391 & 107.469 & 110.150 & 109.311\\
(105) & $\rm BIC$ & 109.185 & 112.767 & 117.388 & 118.085 & 126.074 & 125.235\\
 & $\rm \Delta \ln(BE)$ & - & -1.11 & -0.91 & +0.54 & +0.98 & +0.94\\
\hline
 & $\chi^{2}_{min}$ & 162.508 & 161.267 & 161.196 & 157.332 & 153.754 & 155.512\\
$\rm SN_{sel}$ & $\rm AIC$ & 164.508 & 165.267 & 167.196 & 165.332 & 165.754 & 167.512\\
(182) & $\rm BIC$ & 167.713 & 171.675 & 176.808 & 178.148 & 184.978 & 186.737\\
 & $\rm \Delta \ln(BE)$ & - & -2.23 & -2.70 & -0.50 & +0.52 & +0.40\\
\hline
 & $\chi^{2}_{min}$ & 181.341 & 180.029 & 179.988 & 176.405 & 174.618 & 173.466\\
$\rm SN_{sel}+GRB$ & $\rm AIC$ & 183.341 & 184.029 & 185.988 & 184.405 & 186.618 & 185.466\\
(209) & $\rm BIC$ & 186.683 & 190.713 & 196.015 & 197.774 & 206.672 & 205.520\\
 & $\rm \Delta \ln(BE)$ & - & -2.35 & -2.58 & -0.31 & +0.48 & +0.17\\
\end{tabular}
\end{ruledtabular}
\end{table*}

\begin{table*}[htbp]
\caption{Non-flat case}
\begin{ruledtabular}
\begin{tabular}[t]{c||c|c|c|c|c|c|c}
\bf Data & \bf Type & $\rm \Lambda CDM$ & $\rm XCDM1$ & $\rm XCDM2$ & $\rm X_1X_2CDM$1 & $\rm X_1X_2CDM2$ & $\rm X_1X_2X_3CDM$\\
\hline
 & $\chi^{2}_{min}$ & 156.448 & 156.447 & 156.206 & 156.354 & 155.303 & 156.434\\
$\rm Gold06$ & $\rm AIC$ & 160.448 & 162.447 & 164.206 & 166.354 & 169.303 & 170.434\\
(182) & $\rm BIC$ & 166.856 & 172.059 & 177.022 & 182.374 & 191.731 & 192.863\\
 & $\rm \Delta \ln(BE)$ & - & -1.45 & -1.15 & -1.66 & -1.07 & -1.61\\
\hline
 & $\chi^{2}_{min}$ & 111.006 & 111.066 & 110.495 & 110.641 & 109.711 & 109.544\\
$\rm SNLS$ & $\rm AIC$ & 119.006 & 121.066 & 122.495 & 124.641 & 127.711 & 127.544\\
(115) & $\rm BIC$ & 129.986 & 134.790 & 138.965 & 143.855 & 152.416 & 152.248\\
 & $\rm \Delta \ln(BE)$ & - & -0.80 & -0.07 & -0.78 & +0.46 & -0.14\\
\hline
 & $\chi^{2}_{min}$ & 103.591 & 99.443 & 99.393 & 98.464 & 98.369 & 97.303\\
$\rm ESSENCE$ & $\rm AIC$ & 107.591 & 105.443 & 107.393 & 108.464 & 112.369 & 111.303\\
(105) & $\rm BIC$ & 112.899 & 113.405 & 118.008 & 121.734 & 130.947 & 129.881\\
 & $\rm \Delta \ln(BE)$ & - & -1.10 & +0.29 & -1.07 & -0.14 & -1.30\\
\hline
 & $\chi^{2}_{min}$ & 161.225 & 158.686 & 153.607 & 157.693 & 153.932 & 155.108\\
$\rm SN_{sel}$ & $\rm AIC$ & 165.225 & 164.686 & 161.607 & 167.693 & 167.932 & 169.108\\
(182) & $\rm BIC$ & 171.633 & 174.298 & 174.423 & 183.713 & 185.156 & 191.536\\
 & $\rm \Delta \ln(BE)$ & - & -1.10 & +0.90 & -1.34 & -0.24 & -0.40\\
\hline
 & $\chi^{2}_{min}$ & 180.397 & 178.631 & 178.605 & 176.608 & 174.446 & 173.463\\
$\rm SN_{sel}+GRB$ & $\rm AIC$ & 184.397 & 184.631 & 186.605 & 186.608 & 188.446 & 187.463\\
(209) & $\rm BIC$ & 191.082 & 194.658 & 199.975 & 203.319 & 211.843 & 210.859\\
 & $\rm \Delta \ln(BE)$ & - & -1.32 & -0.11 & -1.27 & +0.08 & +0.03\\
\end{tabular}
\end{ruledtabular}
\end{table*}

\newcommand\AL[3]{~Astron. Lett.{\bf ~#1}, #2~ (#3)}
\newcommand\AP[3]{~Astropart. Phys.{\bf ~#1}, #2~ (#3)}
\newcommand\AJ[3]{~Astron. J.{\bf ~#1}, #2~(#3)}
\newcommand\APJ[3]{~Astrophys. J.{\bf ~#1}, #2~ (#3)}
\newcommand\APJL[3]{~Astrophys. J. Lett. {\bf ~#1}, L#2~(#3)}
\newcommand\APJS[3]{~Astrophys. J. Suppl. Ser.{\bf ~#1}, #2~(#3)}
\newcommand\JCAP[3]{~JCAP. {\bf ~#1}, #2~ (#3)}
\newcommand\LRR[3]{~Living Rev. Relativity. {\bf ~#1}, #2~ (#3)}
\newcommand\MNRAS[3]{~Mon. Not. R. Astron. Soc.{\bf ~#1}, #2~(#3)}
\newcommand\MNRASL[3]{~Mon. Not. R. Astron. Soc.{\bf ~#1}, L#2~(#3)}
\newcommand\NPB[3]{~Nucl. Phys. B{\bf ~#1}, #2~(#3)}
\newcommand\PLB[3]{~Phys. Lett. B{\bf ~#1}, #2~(#3)}
\newcommand\PRL[3]{~Phys. Rev. Lett.{\bf ~#1}, #2~(#3)}
\newcommand\PR[3]{~Phys. Rep.{\bf ~#1}, #2~(#3)}
\newcommand\PRD[3]{~Phys. Rev. D{\bf ~#1}, #2~(#3)}
\newcommand\SJNP[3]{~Sov. J. Nucl. Phys.{\bf ~#1}, #2~(#3)}
\newcommand\ZPC[3]{~Z. Phys. C{\bf ~#1}, #2~(#3)}

\end{document}